\documentclass[prl,epsfigure,twocolumn,10pt,superscriptaddress]{revtex4-2}

\usepackage[dvipsnames]{xcolor} 
\usepackage[utf8]{inputenc}
\usepackage[english]{babel}
\usepackage{amsmath, amsfonts, amssymb}
\usepackage{mathtools}
\usepackage[caption = false]{subfig}
\usepackage{graphicx,epstopdf}
\usepackage{blindtext}
\usepackage{bbm}
\usepackage{enumerate}
\usepackage{multirow}
\usepackage{physics}
\usepackage{times,txfonts}
\usepackage{siunitx}
\usepackage{tabularx}
\usepackage{tikz}
\usepackage[colorlinks=true,linkcolor=blue,urlcolor=blue,citecolor=blue,pdfusetitle]{hyperref}
\usepackage{nicefrac} 

\newcommand\clearrow{\global\let\rowmac\relax}
\clearrow

\allowdisplaybreaks

\usepackage{tikz}
\usetikzlibrary{shapes.geometric, arrows}
\tikzstyle{process} = [rectangle, minimum width=3cm, minimum height=1cm, text centered, draw=white,fill=white]
\tikzstyle{arrow} = [thick,->,>=stealth]

\makeatletter
    \renewcommand\@make@capt@title[2]{%
     \@ifx@empty\float@link{\@firstofone}{\expandafter\href\expandafter{\float@link}}%
      {\textrm{#1}}\@caption@fignum@sep#2\quad}%
    \makeatother
   
\makeatletter 
\renewcommand{\fnum@figure}{\textrm{FIG.~\thefigure}}
\makeatother

\newcommand{\SU}{Department of Physics, Stockholm University, SE-106 91 Stockholm,
Sweden}

\usepackage{orcidlink}

\begin{document}
\title{Rydberg state engineering of trapped ions}

\author{Robin~Thomm~\orcidlink{0009-0000-8105-8690}}
\email{robin.thomm@fysik.su.se}
\affiliation{\SU}

\author{Vinay Shankar~\orcidlink{0009-0005-6551-7689}}
\affiliation{\SU}

\author{Natalia Kuk~\orcidlink{0009-0003-0601-0314}}
\affiliation{\SU}

\author{Marion Mallweger~\orcidlink{0009-0003-0601-0314}}
\affiliation{\SU}

\author{Ivo Straka~\orcidlink{0009-0003-0601-0314}}
\affiliation{\SU}

\author{Markus Hennrich~\orcidlink{0000-0003-2955-7980}} 
\email{markus.hennrich@fysik.su.se}
\affiliation{\SU}

\begin{abstract}

\noindent Microwave dressing of Rydberg ions creates tunable eigenstates with controllable polarizability and interaction strength, but coherent navigation between these states has remained elusive. Here, we report on the first demonstration of coherent population transfer between different Rydberg states of a trapped ion. We investigate both microwave-mediated Rabi oscillations between Rydberg $S$ and $P$ states and adiabatic transfer between microwave-dressed Rydberg states. 
Between Rydberg $S$ and $P$ states we achieve a population transfer efficiency of \SI{91.5(5)}{\percent} in a single microwave $\pi$-pulse. Microwave dressing hybridizes the $S$ and $P$ Rydberg states into new eigenstates with tunable polarizability, enabling both noise-resilient zero-polarizability states and maximally interacting states. We demonstrate adiabatic transfer between these zero-polarizability and maximally dressed states, enabling experiments that combine noise-resilient excitation with strong dipole-dipole interactions and Förster resonance control within a single measurement sequence.

\end{abstract}

\maketitle


\section{Introduction}


Rydberg excitation in atomic systems is a powerful tool for quantum simulation \cite{Bloch2012, Labuhn2016, Gross2017, Scholl2021, Euchner2025}, quantum information processing \cite{Jaksch2000, Saffman2010, Levine2018, Evered2023} and quantum sensing \cite{Sedlacek2012, Sedlacek2013, simons2021, Somaweera2025, Feng2025}. While this is an established field for neutral atoms, Rydberg excitation in trapped ions is a relatively novel approach \cite{Muller_2008, Feldker2015, Higgins2017}.

In neutral atoms, the trapping mechanism relies on dipole forces and is therefore inherently state dependent, typically leading to confinement for low-lying levels but de-confinement for Rydberg states, necessitating turning off the trapping fields during Rydberg excitation \cite{Dudin2012, Levine2018, Graham2019}. While magic trapping of both lower-lying levels and Rydberg levels has been demonstrated \cite{Ahlheit2025, Jansohn2025}, these traps are tailored to one specific Rydberg level and do not allow a change in Rydberg state while maintaining confinement, especially not to a state with opposite polarizability. In contrast, trapped Rydberg ions offer deep confinement for all electronic states, including Rydberg states \cite{Muller_2008}. Together with the excellent control over their internal electronic and external motional degrees of freedom and the advantageous scaling of key atomic properties, such as polarizabilities and interaction strengths, they offer a rich and promising platform \cite{Muller_2008, Mokhberi2020}.

These extreme atomic properties of Rydberg ions have been used in a number of proof-of-principle experiments. The large dipole-dipole interaction strength of microwave-dressed Rydberg states has been employed to implement a two-qubit gate with sub-microsecond duration, independent of the ions' motion \cite{Zhang2020}. Furthermore, the large polarizability of Rydberg states has been central in investigating a structural phase transition of a Wigner crystal triggered by Rydberg excitation \cite{mallweger2025}.

However, trapped Rydberg ions also face challenges. Absorption of black-body radiation can lead to double ionization, and the large polarizabilities of the Rydberg states lead to interactions with the trapping potential. The former can be solved by going to cryogenic temperatures \cite{Mokhberi2020}, whereas the latter leads to a modified trapping potential for Rydberg states \cite{Higgins2019}. The state-dependent trapping strength necessitates ground-state cooling for efficient Rydberg excitation, limiting experiments to low ion numbers, where ground-state cooling of all motional modes of the ions can be achieved.

Zero-polarizability Rydberg states have been realized to overcome this challenge and to achieve efficient and noise-resilient Rydberg excitation for non-ground-state cooled ions \cite{pokorny2020, pokorny_thesis}, but in consequence cancel polarization-related effects and reduce the dipole-dipole interaction strength, which severely limits the experiments that can be performed with zero-polarizability Rydberg states compared to experiments employing other Rydberg states.

Furthermore, all trapped Rydberg ion experiments until now have used only a single Rydberg state for a given experiment \cite{Zhang2020, mallweger2025}. However, fast and coherent control of ions within the Rydberg manifold is essential for more sophisticated experiments, where multiple advantages of the Rydberg states are to be combined within a single experimental sequence. A prominent example is a quantum simulation or quantum algorithm with larger ion crystals, where ground-state cooling of all motional modes may become impractical. In such cases, excitation to a zero-polarizability Rydberg state remains efficient. 
Subsequently, the system can be transferred adiabatically to the target state with the properties required for the application at hand. Depending on the protocol, this may correspond either to a Rydberg state with maximal interaction strength or to a state with a specific polarizability.

Here we demonstrate for the first time such fast and coherent population transfer between bare and dressed Rydberg states of a trapped ion.



\section{Experimental setup}
\begin{figure}[t!]
    \centering
    \includegraphics[width=\columnwidth]{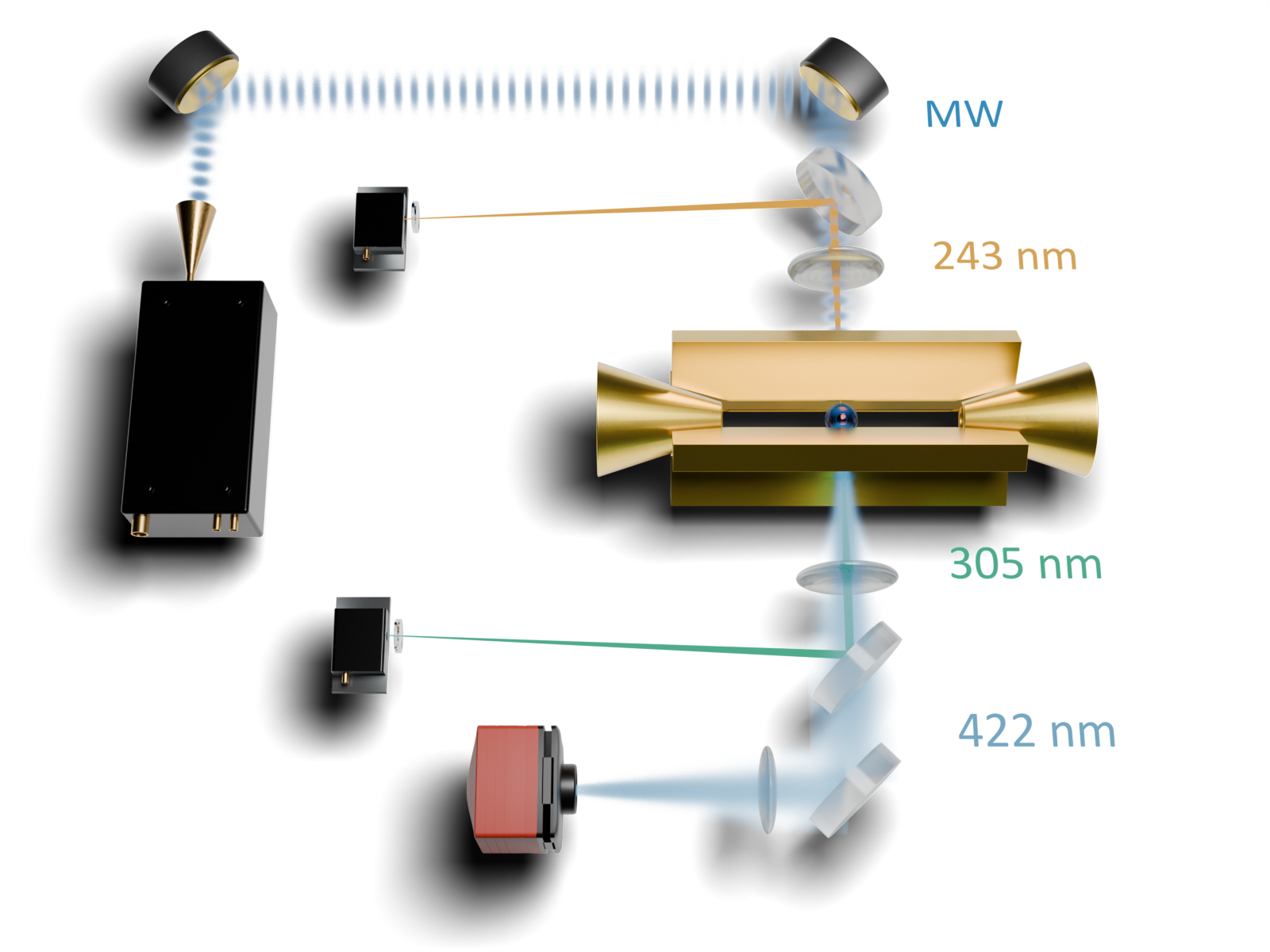}
    
    \vspace{-0.2em}
    \caption{Sketch of the experimental setup. Shown are all lasers needed for Rydberg excitation and manipulation as well as the imaging path. The first Rydberg excitation laser (\SI{243}{\nano\meter}) is overlapped with the MW radiation, the second one (\SI{305}{\nano\meter}) addresses the ion antiparallel to the first one to reduce Doppler broadening of the two-photon resonance and is overlapped with the imaging path (\SI{422}{\nano\meter}). Not shown are the qubit, Doppler cooling, and repumping lasers, which address the ion from the bottom. Furthermore, fluorescence light is collected from above and detected with a photomultiplier tube (PMT).}
    \label{fig:setupMW}
\end{figure}

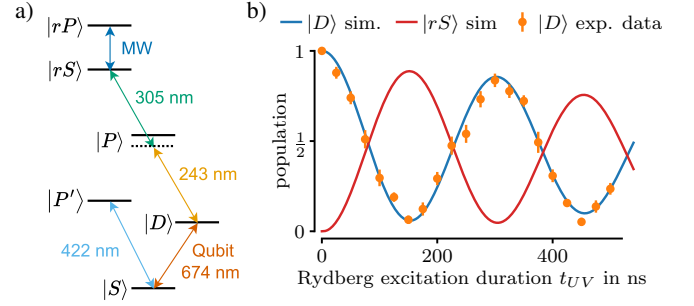
\begin{figure}[!t]
    \centering
    \includegraphics[width=0.35\columnwidth]{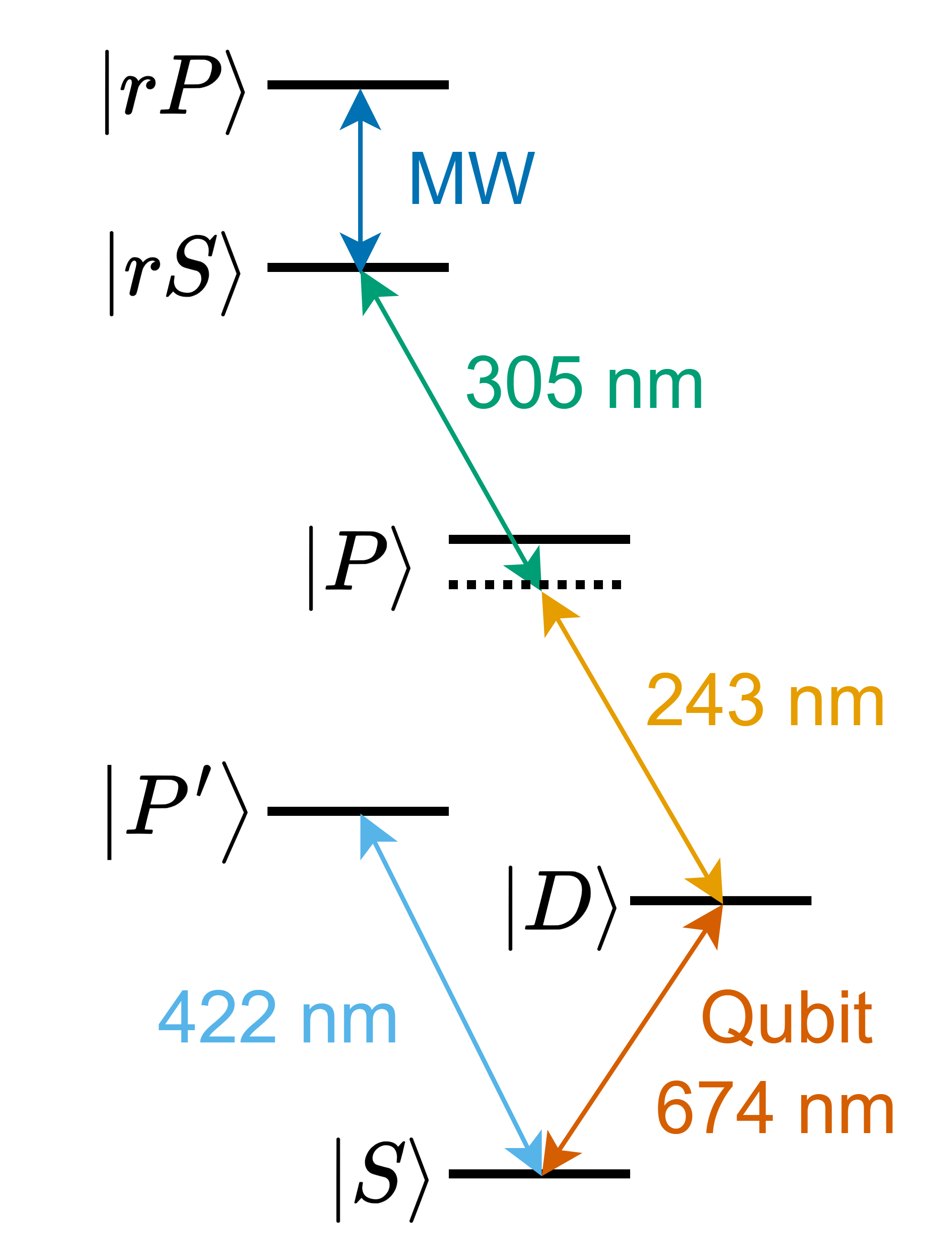}
    \hspace{-0.02\columnwidth}
    \includegraphics[width=0.65\columnwidth]{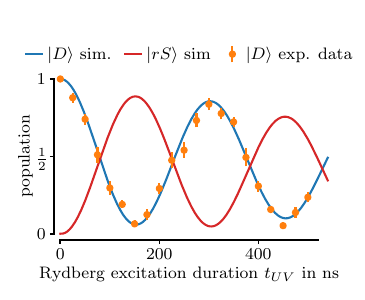}

    \begin{picture}(0,0) 
        \put(-125,120){a)}
    \end{picture}
    \begin{picture}(0,0)
        \put(-40,120){b)}       
    \end{picture}

    \vspace{-1.5em}
    
    \caption{a) Simplified level scheme (see the main text for details), with the main transitions indicated by arrows (repump transitions from metastable states are not shown for clarity). All shown excited ${S}$ and ${P}$ states have a lifetime of below \SI{100}{\micro\second} and ultimately decay to the ground state ${S}$ (or a metastable state that is repumped to $S$). Excitation to one of these states followed by decay can therefore be distinguished from population remaining in the metastable state ${D}$ by fluorescence detection. b) Rabi oscillations between the ${D}$ and ${rS}$ state for \mbox{$n=46$}. We measure the population in ${D}$ (detected as no fluorescence = dark) after a two-photon Rydberg excitation pulse; population in other states decays to ${S}$ and is detected as bright. A simulation (see the Appendix) lets us deduce the $rS$ population (red line) after the Rydberg excitation pulse.}
    \label{fig:levels_and_UV_Rabi}
\end{figure}

For the experiments presented, we used a single $^{88}$Sr$^{+}$ ion trapped in a macroscopic linear Paul trap  \cite{Higgins2017, Higgins2017-2}. A static magnetic field of \SI{0.3}{\milli\tesla} perpendicular to the trap's symmetry axis acts as the quantization axis and lifts the degeneracy of the Zeeman sublevels. A level scheme with all relevant levels is presented in Figure~\ref{fig:levels_and_UV_Rabi} a). The ion is initialized in its electronic ($S = {5S_{1/2}, m_j = -\nicefrac{1}{2}}$) and motional ground-state using standard techniques \cite{Leibfried:03}. We use the ${S} \leftrightarrow {P'} = {5P_{1/2}}$ transition for fluorescence detection and Doppler cooling. The narrow ${S} \leftrightarrow {D} = {4D_{5/2}, m_j = -\nicefrac{5}{2}}$ transition can be coupled with a \SI{674}{\nano\meter} laser, which is also used for sideband cooling.

Rydberg excitation is achieved from the state ${D}$ via a two-photon process \cite{Higgins2021}, where both lasers are switched simultaneously by acousto-optical modulators. The first Rydberg excitation laser with a wavelength of \SI{243}{\nano\meter} couples to an intermediate state ${P} = {6P_{3/2}, m_j = -\nicefrac{3}{2}}$. The second Rydberg excitation laser with a wavelength of \SI{305}{\nano\meter} couples the intermediate state ${P}$ to the Rydberg state \mbox{${rS} = {nS_{1/2}, m_j = -\nicefrac{1}{2}}$}. The Rydberg lasers are aligned radially, (anti-)parallel to the magnetic field, in a counter-propagating configuration to reduce momentum transfer and Doppler broadening during the excitation process. We use $\sigma^+$ polarized light to only address the Zeeman sublevels specified and to minimize off-resonant coupling to other Zeeman sublevels.

To avoid scattering from the short-lived intermediate state ${P}$, the Rydberg lasers are detuned by $\Delta_P = -\SIrange{120}{200}{\mega\hertz}$ depending on the experiment and the available laser power. The effective Rabi frequency between ${S}$ and ${rS}$ of the Rydberg excitation is given by
\begin{equation}
    \Omega_\mathrm{R} = \frac{\Omega_{243}\Omega_{305}}{2 \Delta_P} \, ,
\end{equation}
with Rabi frequencies $\Omega_{243}$ and $\Omega_{305}$ of the two individual transitions, respectively. Within the Rydberg manifold, we couple the ${rS}$ state with the Rydberg state \mbox{${rP} = {nP_{1/2}, m_j=\nicefrac{1}{2}}$} via microwave (MW) radiation.

In the following experiments we use principal quantum numbers of $n=46$ for the MW Rabi oscillations and $n=56$ for the adiabatic transfer between different MW-dressed Rydberg states. We choose $n=46$ as a tradeoff where high Rydberg state polarizabilities and interaction strengths are present but still low enough that the coupling to the trapping potential is small. To be able to realize adiabatic transfer, we replace a frequency doubler in the MW generation setup (see the Appendix for a detailed description) with a mixer, which halves the frequency range we can address with the MW radiation, making it necessary to employ higher Rydberg states with smaller splitting between Rydberg $rS$ and $rP$ states.

The MW power is calibrated with an indirect measurement as done in \cite{Higgins2021, Mokhberi2020, bao2025}. The ${rS} \leftrightarrow {rP}$ transition is driven on resonance, creating dressed states and splitting the ${rS}$ level by the MW Rabi frequency $\Omega_\mathrm{MW}$ due to the Autler-Townes effect. The ${P} \leftrightarrow {rS}$ transition is then probed with the \SI{305}{\nano\meter} laser. If the \SI{305}{\nano\meter} laser is resonant to one of the dressed states, it will in turn couple the dressed state to the ${P}$ state, creating an Autler-Townes splitting of the ${P}$ level. The \SI{243}{\nano\meter} laser, resonant on the ${D} \leftrightarrow {P}$ transition, will not be on resonance anymore and the population stays in ${D}$. On the other hand, when the \SI{305}{\nano\meter} laser is not resonant to one of the dressed states, it will not lead to a splitting of the ${P}$ level so that the \SI{243}{\nano\meter} laser is on resonance and the population in the ${D}$ state is pumped to the ${S}$ state via the short-lived ${P}$ state. Using this technique, we measure a MW Rabi frequency of \mbox{$2\pi \times \SI{12.3(5)}{\mega\hertz}$} for the principal quantum number $n=46$ for the MW power used in the subsequent experiments.

\begin{figure*}[!t]
    \centering
    \includegraphics[width=0.78\columnwidth]{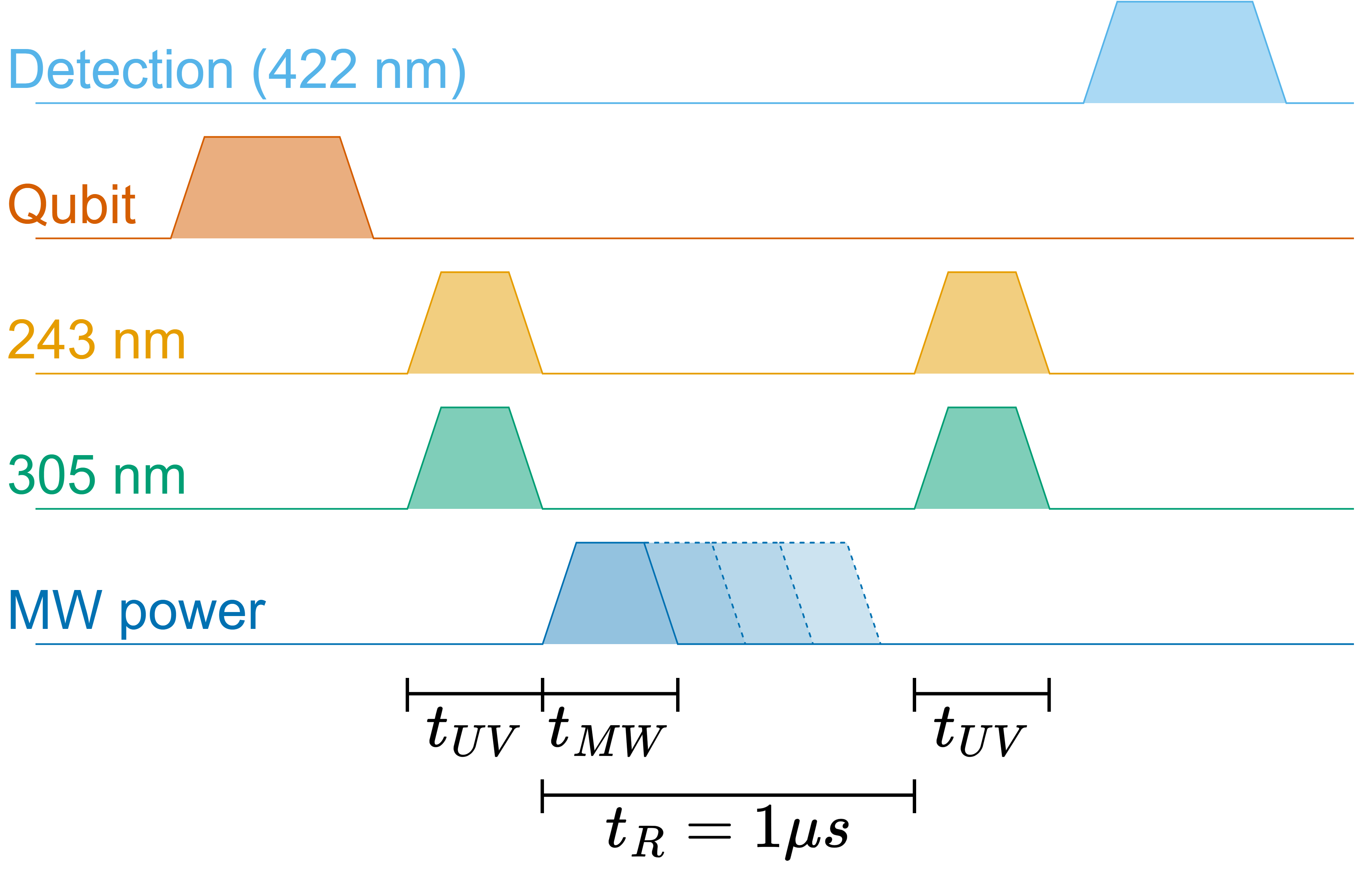}
    \hspace{10pt}
    \includegraphics{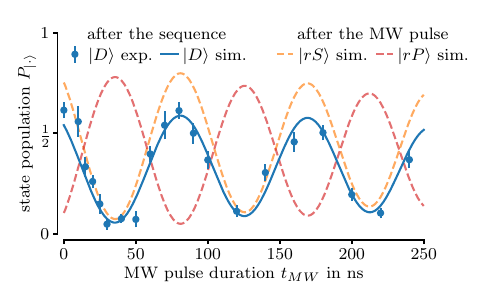}

    \begin{picture}(0,0) 
        \put(-230,135){a)}
    \end{picture}
    \begin{picture}(0,0) 
        \put(-15,135){b)}       
    \end{picture}

    \vspace{-2.em}
    \caption{a) Pulse sequence used for exciting and manipulating the Rydberg state. We start with an ion initialized in its electronic and motional ground state. A qubit $\pi$-pulse excites it into the metastable ${D}$ state. From there we excite the ion via a two-photon process to the Rydberg state (cf.~Fig.~\ref{fig:levels_and_UV_Rabi}). Once in the Rydberg manifold, we coherently transfer population between different Rydberg states via MW radiation. We probe the success of the coherent manipulation by driving the fraction remaining in the Rydberg ${rS}$ state back to ${D}$ with a two-photon de-excitation pulse, all other population decays to ${S}$. The experiment is concluded with a fluorescence measurement. b) MW Rabi oscillations recorded with the pulse sequence from a). The points show the experimentally measured ${D}$ state population, the blue line the simulated result. The dashed lines show the simulated Rydberg ${rS}$ and ${rP}$ state populations directly after the MW pulse. We achieve a population transfer efficiency with a MW $\pi$-pulse of \SI{91.5(5)}{\percent}.}
    \label{fig:pulse_sequence_MWrabi}
\end{figure*}
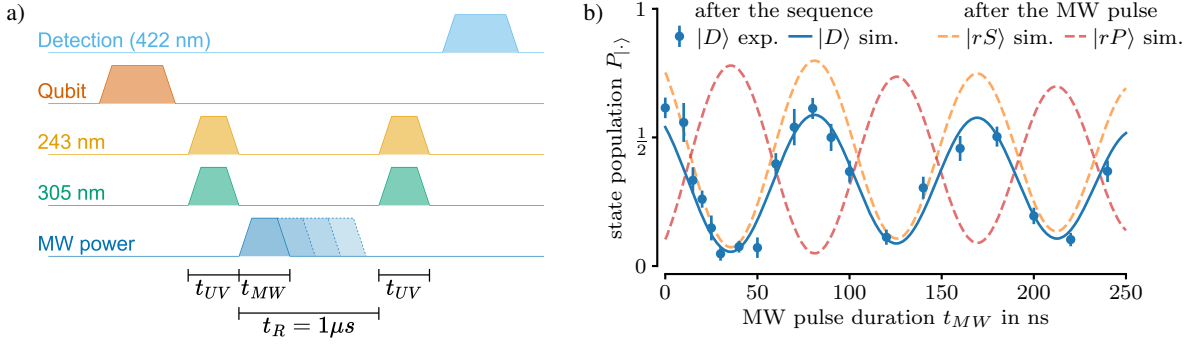

\section{Microwave Rabi oscillations}\label{section:results}
To study coherent population transfer between the bare Rydberg states ${rS}$ and ${rP}$, we employ a principal quantum number of $n=46$ and drive the transition resonantly with MW radiation of about \SI{120}{\giga\hertz}. The MW is switched between the primary frequency source of \SI{15}{\giga\hertz} and the frequency multiplication steps, which consist of a quadrupling and amplifying stage, followed by a second frequency doubling stage. The MW is then coupled to free space via a conical horn antenna and collimated and focused onto the ion from the radial direction by two parabolic gold-coated mirrors (see the Appendix for a more detailed description of the MW setup and calibration, and Fig.~\ref{fig:setupMW} for a sketch of the delivery of the MW to the ion).


In a first step, we probe the Rydberg excitation to maximize the excitation efficiency. We use standard techniques of Doppler cooling and resolved sideband cooling \cite{Leibfried:03} to achieve a thermal excitation of the radial motional modes of $\bar{n} < 0.1$\,. 
We then excite the ion from state $D$ to Rydberg state $rS$ by switching both Rydberg lasers simultaneously. We keep the two-photon transition on resonance while detuning the Rydberg lasers from the intermediate state ${P}$ by \SI{-200}{\mega\hertz}. We fine-tune the resonance frequency by probing the two-photon transition with the \SI{243}{\nano\meter} laser and probe the coupling strength by varying the length of the pulse. In this way we drive Rabi oscillations between ${D}$ and ${rS}$, presented in Figure~\ref{fig:levels_and_UV_Rabi} b). From the Rabi oscillations we extract the effective Rabi frequency of $\Omega_\mathrm{R} = 2\pi \times \SI{3.334(2)}{\mega\hertz}$ and a Rydberg excitation efficiency of~$\SI{89}{\percent}$.

To probe Rabi oscillations between the Rydberg states, we perform two Rydberg excitation $\pi$-pulses after the excitation to ${D}$, separated by a wait time of $t_\mathrm{R} = \SI{1}{\micro\second}$ where the ion resides in the Rydberg manifold. During this time we apply a MW pulse of variable time $t_\mathrm{MW}$. This MW pulse drives Rabi oscillations between the ${rS}$ and ${rP}$ states. All the population remaining in ${rS}$ after the MW pulse and the wait time is transferred back to ${D}$ by the second Rydberg de-excitation $\pi$-pulse and detected as a dark measurement, whereas population in ${rP}$ decays to ${S}$ in a multi-step process and is detected as a bright measurement. The pulse sequence is shown in Figure~\ref{fig:pulse_sequence_MWrabi} a).

Figure~\ref{fig:pulse_sequence_MWrabi} b) shows the experimental results together with the simulated expectation; error bars correspond to quantum projection noise and depict $1\sigma$ confidence intervals. A clear oscillation in ${D}$ state population can be seen with a frequency of $\Omega_\mathrm{MW} = 2\pi \times \SI{11.1(3)}{\mega\hertz}$, within the $2\sigma$ confidence interval of the expected value of $2\pi \times \SI{12.3(5)}{\mega\hertz}$. The low maximum population in ${D}$ is due to the relatively short Rydberg state lifetime of the ${rS}$ level of $\tau_{rS} = \SI{4.6}{\micro\second}$, which leads to decay of a large fraction of the ${rS}$ state to ${S}$ while the ion is in the Rydberg state, and the limited efficiency of the Rydberg excitation pulse. The observed dynamics are very close to the evolution expected from the simulation (solid line, see the Appendix for details on the simulation). From the simulation we extract the populations of the Rydberg states ${rS}$ and ${rP}$ directly after the MW pulse (dashed lines). \SI{91.5(5)}{\percent} of the Rydberg state population is transferred to ${rP}$ after a MW $\pi$-pulse.

\section{Adiabatic transfer between dressed Rydberg states}
\begin{figure*}
    \centering
    \includegraphics[width=1.8\columnwidth]{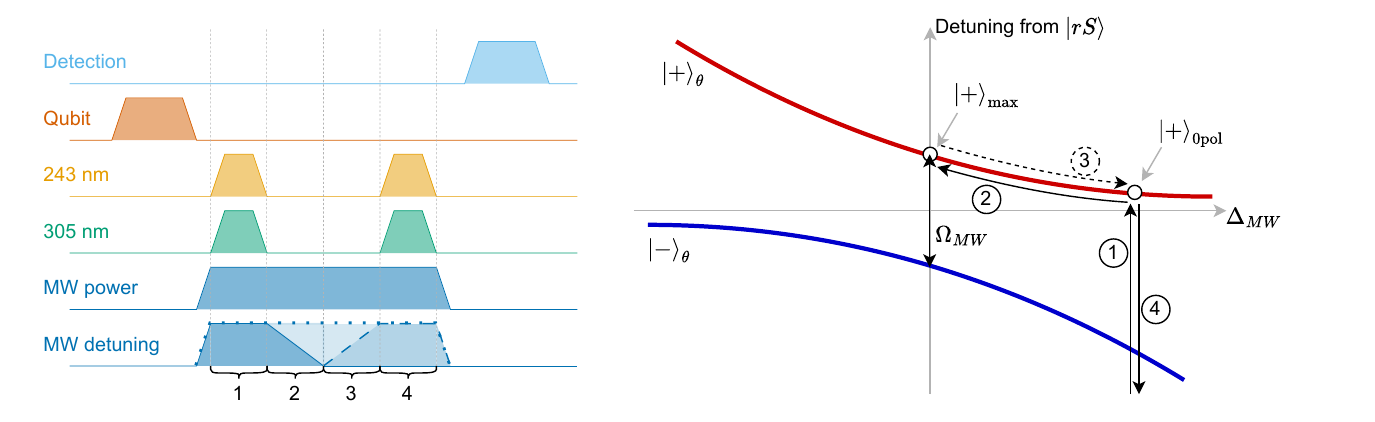}

    \begin{picture}(0,0) 
        \put(-220,130){a)}
    \end{picture}
    \begin{picture}(0,0) 
        \put(-20,130){b)}       
    \end{picture}

    \vspace{-1.5em}
    \caption{Experimental sequence for probing the adiabatic transfer between different MW-dressed Rydberg states. In a) the pulse sequence is shown, in b) the dressed state energies dependent on the MW detuning $\Delta_\mathrm{MW}$. We again start by transferring population to ${D}$. The Rydberg excitation and manipulation is then divided into four periods. In periods 1 and 4 both Rydberg lasers are switched simultaneously to transfer population between ${D}$ and $\ket{+}_\mathrm{0pol}$. In periods 2 and 3 the MW detuning can be swept (see the text for a more detailed description of the periods). During the whole time of periods 1 to 4, the MW irradiates the ion with variable detunings. We investigate 3 different cases, a constant detuning (dotted), sweeping the MW detuning to \SI{0}{\mega\hertz} in period 2 and back to \SI{50}{\mega\hertz} in period 3 (dashed), and sweeping the MW detuning to \SI{0}{\mega\hertz} in period 2, but keeping it there for periods 3 and 4 (straight line).}
    \label{fig:scheme_MWsweep}
\end{figure*}
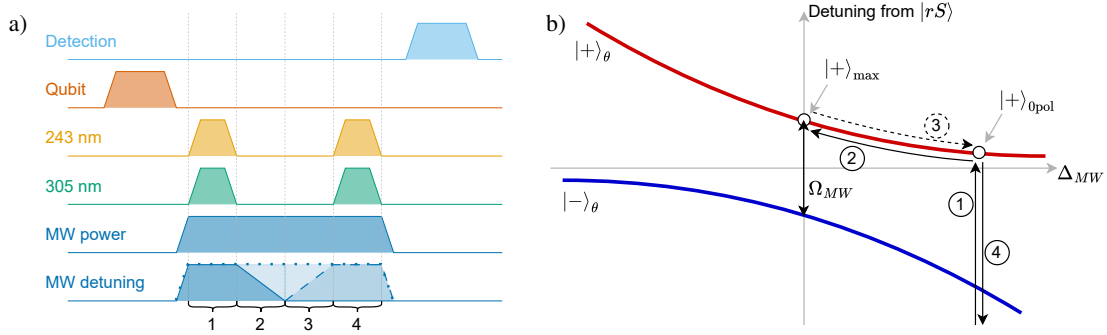

After demonstrating coherent population transfer between bare Rydberg states, the logical next step is to investigate transfer between MW-dressed Rydberg states. Here, the Rydberg states ${rS}$ and ${rP}$ with a principal quantum number of $n=56$ are coupled with MW radiation of about \SI{65.5}{\giga\hertz}.

The coupling of the MW leads to state hybridization and creates new eigenstates \cite{pokorny2020, pokorny_thesis}, also known as dressed states
\begin{align}
    \ket{+} = \sin(\theta)\ket{rS} + \cos(\theta)\ket{rP} \, , \nonumber\\
    \ket{-} = \cos(\theta)\ket{rS} + \sin(\theta)\ket{rP} \, ,
\end{align}
with mixing angle
\begin{equation}
    \tan(\theta) = \frac{\Omega_\mathrm{MW}}{\sqrt{\Omega_\mathrm{MW}^2 + \Delta_\mathrm{MW}^2} - \Delta_\mathrm{MW}} \, ,
\end{equation}
see also Figure \ref{fig:scheme_MWsweep} (right). Due to the higher Rydberg state employed in comparison to the measurements described above, the interaction of the quadrupolar electric trapping field to the Rydberg states, in particular the ${rP}$ state, becomes much stronger \cite{Higgins2021}. To suppress this effect, we operate in a weaker trap with radial motional frequencies of \mbox{$(\omega_x, \omega_y) = 2\pi \times (0.4, 0.6)$ \si{\mega\hertz}}, compared to \mbox{$(\omega_x, \omega_y) = 2\pi \times (1.4, 1.5)$ \si{\mega\hertz}} for the measurements presented above. However, this also reduces the effectivity of the cooling to the motional ground state, leading to a higher initial phonon number of $\bar{n}_x + \bar{n}_y \approx 5$ in the radial motional modes.

Since the ${rS}$ and ${rP}$ states have polarizabilities with opposite signs $\mathcal{P}_{{rS}} > 0 > \mathcal{P}_{{rP}}$, we can achieve dressed states with zero effective polarizability, making them insensitive to first-order effects of stray electric fields, shifts in the trapping potential and line broadening effects due to imperfect ground-state cooling \cite{Higgins2019, Higgins2017, pokorny2020, pokorny_thesis}. For the $\ket{+}$ dressed states, this zero-polarizability state $\ket{+}_\mathrm{0pol}$ is characterized by a mixing angle of
\begin{equation}
    \tan(\theta_\mathrm{0pol}) = \sqrt{- \frac{\mathcal{P}_{{rP}}}{\mathcal{P}_{{rS}}}} \, .
\end{equation}
With a MW Rabi frequency of $\Omega_\mathrm{MW} = 2\pi \times \SI{65.5}{\mega\hertz}$ we achieve the zero-polarizability state $\ket{+}_{\mathrm{0pol}}$ with a MW detuning of $\Delta_\mathrm{MW} \approx 2\pi \times \SI{50}{\mega\hertz}$ and a corresponding detuning of $\Delta_\mathrm{R} = 2\pi \times \SI{16}{\mega\hertz}$ of $\ket{+}_{\mathrm{0pol}}$ from ${rS}$.

\begin{table}[b!]
    \centering

    \begin{tabular}{crrrr}
        \multirow{ 2}{*}{case } & \multirow{ 2}{*}{ $\Delta_\mathrm{MW}$ in period 2 } & \multirow{ 2}{*}{ $\Delta_\mathrm{MW}$ in period 3 } & \multicolumn{2}{c}{return probability} \\
         &  &  & simulated & measured \\
        \hline
        1 & constant \SI{50}{\mega\hertz} & constant \SI{50}{\mega\hertz} & \SI{53}{\percent} & \SI{50(6)}{\percent}\\
        2 & $(50 \searrow 0)$ \si{\mega\hertz} & $(0 \nearrow 50)$ \si{\mega\hertz} & \SI{54}{\percent} & \SI{42(4)}{\percent}\\
        3 & $(50 \searrow 0)$ \si{\mega\hertz} & constant \SI{0}{\mega\hertz} & \SI{4}{\percent} & \SI{8(4)}{\percent}\\
    \end{tabular}
    \vspace{-0.2em}
    \caption{Summary of the MW sweeping results. The simulation was done similarly to the one in Figure~\ref{fig:pulse_sequence_MWrabi}, see also the Appendix. We have good agreement with the simulation for case 1 (no sweeping), and lower efficiency for case 2 (sweeping both ways). For case 3, we again have good agreement with the simulated result, indicating that we fully remove the population from the zero polarizability state during the MW sweep.}
    \label{tab:results_MWsweeping}
\end{table}

We investigate the adiabatic transfer from $\ket{+}_{\mathrm{0pol}}$ to the maximally dressed state $\ket{+}_{\mathrm{max}}$ with $\Delta_\mathrm{MW} = 0$. 
The Rydberg excitation and manipulation is divided into 4 periods and assumes that the ion is initially prepared in its electronic ${D}$ state (see also Fig.~\ref{fig:scheme_MWsweep}). During the whole time, the ion is irradiated with the MW, initially detuned by $\Delta_\mathrm{MW} \approx 2\pi \times \SI{50}{\mega\hertz}$ to create $\ket{+}_{\mathrm{0pol}}$. During the first period, the Rydberg lasers are switched on simultaneously for \SI{200}{\nano\second} to transfer population from ${D}$ to $\ket{+}_{\mathrm{0pol}}$. We achieve a population transfer of \SI{88}{\percent}. 
During the second and third periods, taking \SI{500}{\nano\second} each, the MW is swept linearly in frequency (see also Fig.~\ref{fig:scheme_MWsweep} and Tab.~\ref{tab:results_MWsweeping}). During these sweeps, the ion can be adiabatically transferred between the zero-polarizability state $\ket{+}_\mathrm{0pol}$ and the maximally dressed state $\ket{+}_\mathrm{max}$. 
In the fourth period, we apply again a simultaneous pulse of the Rydberg lasers identical to the one in period 1, transferring the population from $\ket{+}_{\mathrm{0pol}}$ to ${D}$. After this sequence, we perform fluorescence detection. Again, all population that is not transferred back to ${D}$ will eventually decay to ${S}$ in a multi-step process and is detected as a bright measurement.

We investigate three cases. First, we keep the MW detuning constant during periods 2 and 3, leaving the population to reside in $\ket{+}_{\mathrm{0pol}}$. Second, we sweep the MW frequency from $\Delta_\mathrm{MW} = \SI{50}{\mega\hertz}$ to $\Delta_\mathrm{MW} = \SI{0}{\mega\hertz}$ in period 2 and sweep back to $\Delta_\mathrm{MW} = \SI{50}{\mega\hertz}$ in period 3. Here, we expect the ion to adiabatically follow the eigenstate from $\ket{+}_{\mathrm{0pol}}$ to $\ket{+}_{\mathrm{max}}$ and back to $\ket{+}_{\mathrm{0pol}}$. The third case is similar to the second, but the sweeping back is skipped and the MW detuning is kept at zero during periods 3 and 4. Here, we expect the population to stay in $\ket{+}_{\mathrm{max}}$, making the Rydberg laser pulse in period 4 ineffective since it is off-resonant to $\ket{+}_{\mathrm{max}}$ and $\Omega_\mathrm{R} \ll \abs{\Delta_{\mathrm{R}, \mathrm{0pol}} - \Delta_{\mathrm{R}, \mathrm{max}}}$.

We measure return probabilities from the Rydberg state to state ${D}$ for the three cases of \SI{50(6)}{\percent}, \SI{42(4)}{\percent} and \SI{8(4)}{\percent}, see also Table~\ref{tab:results_MWsweeping} for a comparison with simulated values.

For case 1, we see good agreement with the expected return probability, as is expected since no sweeping is performed. The lower-than-expected return probability for case 2 indicates some loss mechanism during the sweep, which is not represented in the simulation. Possible loss mechanisms include a too low estimate of the Rydberg state dephasing, non-adiabatic processes due to finite frequency steps during the MW sweep, and coupling to other levels, potentially assisted by the RF trapping potential during the sweep or a \mbox{$\pi$-polarization} component in the MW radiation, stemming from imperfect alignment of the MW with respect to the magnetic field. This $\pi$-polarization component would then couple to the ${56P_{1/2}, m_j=-\nicefrac{1}{2}}$ state. However, the small return probability in case 3 verifies that most of the population leaves $\ket{+}_{\mathrm{0pol}}$ during the sweep in period 2.

The dominant source of low return probability in all three cases is, as before, the Rydberg state lifetime at room temperature (\SI{300}{\kelvin}) of \SI{8.2}{\micro\second} of the ${rS}$ state and double ionization due to absorption of blackbody radiation photons. The latter is especially notable for the ${rP}$ state with a lifetime of \SI{40}{\micro\second} at room temperature, and limits the number of repetitions due to the necessity of frequent ion reloading.




\section{Conclusion}
In this work, we demonstrated Rabi oscillations between two distinct Rydberg states on sub-microsecond timescales with an efficiency of \SI{91.5(5)}{\percent}, and adiabatic population transfer between different MW-dressed Rydberg states. Together, these results establish a flexible toolbox for experiments that incorporate multiple Rydberg states within a single measurement sequence.


The demonstrated fast adiabatic transfer enables excitation to arbitrary MW-dressed Rydberg states via the zero polarizability state \cite{pokorny2020, pokorny_thesis}. This allows efficient excitation in the presence of electric field noise and from non-ground-state cooled ions, followed by adiabatic transfer to any desired dressed Rydberg state. The maximally dressed state is a natural candidate owing to its maximum dipole-dipole interaction strength \cite{Zhang2020}, while states with predominantly Rydberg $P$ population offer longer lifetimes and opposite sign of polarizability, enabling control over the mechanical forces acting on the Rydberg ions. Crucially, the techniques demonstrated here give coherent access to both within a single experimental sequence. With these methods, future experiments with longer ion strings can exploit the large polarizabilities of Rydberg ions, which was previously unattainable.

Furthermore, pulsed application of the MW allows the preparation of different ions in different Rydberg states, potentially unlocking Förster resonances and the associated energy transfer, a new type of interaction for trapped ions that has seen much attention in neutral atom experiments \cite{Gorniaczyk2016, Nipper2012, Ravets2014}. 
Engineering coherent driving between Rydberg states via MW radiation can also enable coherent coupling to superconducting qubits, opening a route towards hybrid quantum devices \cite{Kaiser2022, Petrosyan2008, Petrosyan2009, Hogan2012}.

Overall, the techniques presented here significantly broaden the range of controllable interactions and accessible Rydberg states in trapped-ion systems. They offer a promising pathway toward more robust Rydberg-ion quantum computing and quantum simulations with larger ion crystals.


\section*{Acknowledgements}
We thank Weibin Li for theoretical data on Rydberg lifetimes, polarizabilities, and decay processes. This work was supported by the Knut \& Alice Wallenberg Foundation (Wallenberg Centre for Quantum Technology [WACQT]), by the Swedish Research Council (Grant No. 2021-05811). This project has also received funding from the European Union’s Horizon Europe research and innovation program under Grant Agreement No. 101046968 (BRISQ). This work was supported by the QuantERA II Programme (project CoQuaDis) that has received funding from the EU H2020 research and innovation programme under GA No. 101017733. This work is supported by the ERC grant OPEN-2QS (Grant No.\ 101164443, https://doi.org/10.3030/101164443).

\bibliography{My_bib.bib}

@article{Kaiser2022,
    title = {Cavity-driven {R}abi oscillations between {R}ydberg states of atoms trapped on a superconducting atom chip},
    author = {Kaiser, Manuel and Glaser, Conny and Ley, Li Yuan and Grimmel, Jens and Hattermann, Helge and Bothner, Daniel and Koelle, Dieter and Kleiner, Reinhold and Petrosyan, David and G\"unther, Andreas and Fort\'agh, J\'ozsef},
    journal = {Phys. Rev. Res.},
    volume = {4},
    issue = {1},
    pages = {013207},
    numpages = {12},
    year = {2022},
    month = {Mar},
    publisher = {American Physical Society},
    doi = {10.1103/PhysRevResearch.4.013207},
    url = {https://link.aps.org/doi/10.1103/PhysRevResearch.4.013207}
}

@article{Leibfried:03,
	title = {Quantum dynamics of single trapped ions},
	author = {Leibfried, D. and Blatt, R. and Monroe, C. and Wineland, D.},
	journal = {Rev. Mod. Phys.},
	volume = {75},
	issue = {1},
	pages = {281--324},
	numpages = {0},
	year = {2003},
	month = {Mar},
	publisher = {American Physical Society},
	doi = {10.1103/RevModPhys.75.281}
}

@article{Feng2025,
    author={Feng, Zhigang
    and Liu, Xiaochi
    and Song, Zhenfei
    and Qu, Jifeng},
    title={Multi-parameter microwave quantum sensing with a single atomic probe},
    journal={Scientific Reports},
    year={2025},
    month={Feb},
    day={13},
    volume={15},
    number={1},
    pages={5379},
    issn={2045-2322},
    doi={10.1038/s41598-025-89697-4},
    url={https://doi.org/10.1038/s41598-025-89697-4}
}

@article{Petrosyan2009,
    title = {Reversible state transfer between superconducting qubits and atomic ensembles},
    author = {Petrosyan, David and Bensky, Guy and Kurizki, Gershon and Mazets, Igor and Majer, Johannes and Schmiedmayer, J\"org},
    journal = {Phys. Rev. A},
    volume = {79},
    issue = {4},
    pages = {040304},
    numpages = {4},
    year = {2009},
    month = {Apr},
    publisher = {American Physical Society},
    doi = {10.1103/PhysRevA.79.040304},
    url = {https://link.aps.org/doi/10.1103/PhysRevA.79.040304}
}

@article{Hogan2012,
    title = {Driving {R}ydberg\-{R}ydberg Transitions from a Coplanar Microwave Waveguide},
    author = {Hogan, S. D. and Agner, J. A. and Merkt, F. and Thiele, T. and Filipp, S. and Wallraff, A.},
    journal = {Phys. Rev. Lett.},
    volume = {108},
    issue = {6},
    pages = {063004},
    numpages = {5},
    year = {2012},
    month = {Feb},
    publisher = {American Physical Society},
    doi = {10.1103/PhysRevLett.108.063004},
    url = {https://link.aps.org/doi/10.1103/PhysRevLett.108.063004}
}

@article{Petrosyan2008,
    title = {Quantum Information Processing with Single Photons and Atomic Ensembles in Microwave Coplanar Waveguide Resonators},
    author = {Petrosyan, David and Fleischhauer, Michael},
    journal = {Phys. Rev. Lett.},
    volume = {100},
    issue = {17},
    pages = {170501},
    numpages = {4},
    year = {2008},
    month = {Apr},
    publisher = {American Physical Society},
    doi = {10.1103/PhysRevLett.100.170501},
    url = {https://link.aps.org/doi/10.1103/PhysRevLett.100.170501}
}

@article{Higgins2017-2,
  title = {Coherent Control of a Single Trapped {R}ydberg Ion},
  author = {Higgins, Gerard and Pokorny, Fabian and Zhang, Chi and Bodart, Quentin and Hennrich, Markus},
  journal = {Phys. Rev. Lett.},
  volume = {119},
  issue = {22},
  pages = {220501},
  numpages = {5},
  year = {2017},
  month = {Nov},
  publisher = {American Physical Society},
  doi = {10.1103/PhysRevLett.119.220501},
  url = {https://link.aps.org/doi/10.1103/PhysRevLett.119.220501}
}

@article{Higgins2017,
  title = {Single Strontium {R}ydberg Ion Confined in a {P}aul Trap},
  author = {Higgins, Gerard and Li, Weibin and Pokorny, Fabian and Zhang, Chi and Kress, Florian and Maier, Christine and Haag, Johannes and Bodart, Quentin and Lesanovsky, Igor and Hennrich, Markus},
  journal = {Phys. Rev. X},
  volume = {7},
  issue = {2},
  pages = {021038},
  numpages = {11},
  year = {2017},
  month = {Jun},
  publisher = {American Physical Society},
  doi = {10.1103/PhysRevX.7.021038},
  url = {https://link.aps.org/doi/10.1103/PhysRevX.7.021038}
}

@article{Feldker2015,
    title = {Rydberg Excitation of a Single Trapped Ion},
    author = {Feldker, T. and Bachor, P. and Stappel, M. and Kolbe, D. and Gerritsma, R. and Walz, J. and Schmidt-Kaler, F.},
    journal = {Phys. Rev. Lett.},
    volume = {115},
    issue = {17},
    pages = {173001},
    numpages = {5},
    year = {2015},
    month = {Oct},
    publisher = {American Physical Society},
    doi = {10.1103/PhysRevLett.115.173001},
    url = {https://link.aps.org/doi/10.1103/PhysRevLett.115.173001}
}

@article{Muller_2008,
    doi = {10.1088/1367-2630/10/9/093009},
    url = {https://doi.org/10.1088/1367-2630/10/9/093009},
    year = {2008},
    month = {sep},
    publisher = {},
    volume = {10},
    number = {9},
    pages = {093009},
    author = {Müller, Markus and Liang, Linmei and Lesanovsky, Igor and Zoller, Peter},
    title = {Trapped {R}ydberg ions: from spin chains to fast quantum gates},
    journal = {New Journal of Physics}
}

@misc{mallweger2025,
      title={Probing electronic state-dependent conformational changes in a trapped {R}ydberg ion {W}igner crystal}, 
      author={Marion Mallweger and Natalia Kuk and Vinay Shankar and Robin Thomm and Harry Parke and Ivo Straka and Weibin Li and Igor Lesanovsky and Markus Hennrich},
      year={2025},
      eprint={2507.23631},
      archivePrefix={arXiv},
      primaryClass={quant-ph},
      url={https://arxiv.org/abs/2507.23631}, 
}

@misc{pokorny2020,
      title={Magic trapping of a {R}ydberg ion with a diminished static polarizability}, 
      author={Fabian Pokorny and Chi Zhang and Gerard Higgins and Markus Hennrich},
      year={2020},
      eprint={2005.12422},
      archivePrefix={arXiv},
      primaryClass={physics.atom-ph},
      url={https://arxiv.org/abs/2005.12422}, 
}

@misc{wilkinson2024,
      title={Two-qubit gate protocols with microwave-dressed Rydberg ions in a linear Paul trap}, 
      author={Joseph W. P. Wilkinson and Katrin Bolsmann and Thiago L. M. Guedes and Markus Müller and Igor Lesanovsky},
      year={2024},
      eprint={2412.13699},
      archivePrefix={arXiv},
      primaryClass={quant-ph},
      url={https://arxiv.org/abs/2412.13699}, 
}

@misc{Weibin,
  author = "Li, Weibin",
  howpublished = "private communication"
}

@article{Kiruga2026,
    title = {Portal for high-precision atomic data and computation},
    journal = {Computer Physics Communications},
    volume = {319},
    pages = {109951},
    year = {2026},
    issn = {0010-4655},
    doi = {https://doi.org/10.1016/j.cpc.2025.109951},
    url = {https://www.sciencedirect.com/science/article/pii/S0010465525004527},
    author = {Amani Kiruga and Charles Cheung and Dmytro Filin and Parinaz Barakhshan and Akshay Bhosale and Vipul Badhan and Bindiya Arora and Rudolf Eigenmann and Marianna S. Safronova},
}

@article{Jansohn2025,
  title = {Magic-wavelength trapping of alkali-metal {R}ydberg atoms: The role of landscape polarization modulation},
  author = {Jansohn, C. and Londo\~no, A. and Li, Y. and Nguyen, H. and Berman, P. R. and Kuzmich, A.},
  journal = {Phys. Rev. A},
  volume = {112},
  issue = {4},
  pages = {L041101},
  numpages = {6},
  year = {2025},
  month = {Oct},
  publisher = {American Physical Society},
  doi = {10.1103/43wm-383m},
  url = {https://link.aps.org/doi/10.1103/43wm-383m}
}

@article{Ahlheit2025,
  title = {Magic running- and standing-wave optical traps for {R}ydberg atoms},
  author = {Ahlheit, Lukas and Nill, Chris and Svirskiy, Daniil and de Haan, Jan and Schroers, Simon and Alt, Wolfgang and Stiesdal, Nina and Lesanovsky, Igor and Hofferberth, Sebastian},
  journal = {Phys. Rev. A},
  volume = {111},
  issue = {1},
  pages = {013115},
  numpages = {10},
  year = {2025},
  month = {Jan},
  publisher = {American Physical Society},
  doi = {10.1103/PhysRevA.111.013115},
  url = {https://link.aps.org/doi/10.1103/PhysRevA.111.013115}
}

@article{Graham2019,
    title = {Rydberg-Mediated Entanglement in a Two-Dimensional Neutral Atom Qubit Array},
    author = {Graham, T. M. and Kwon, M. and Grinkemeyer, B. and Marra, Z. and Jiang, X. and Lichtman, M. T. and Sun, Y. and Ebert, M. and Saffman, M.},
    journal = {Phys. Rev. Lett.},
    volume = {123},
    issue = {23},
    pages = {230501},
    numpages = {6},
    year = {2019},
    month = {Dec},
    publisher = {American Physical Society},
    doi = {10.1103/PhysRevLett.123.230501},
    url = {https://link.aps.org/doi/10.1103/PhysRevLett.123.230501}
}

@article{Dudin2012,
    author={Dudin, Y. O.
    and Li, L.
    and Bariani, F.
    and Kuzmich, A.},
    title={Observation of coherent many-body {R}abi oscillations},
    journal={Nature Physics},
    year={2012},
    month={Nov},
    day={01},
    volume={8},
    number={11},
    pages={790-794},
    issn={1745-2481},
    doi={10.1038/nphys2413},
    url={https://doi.org/10.1038/nphys2413}
}

@article{Jaksch2000,
    title = {Fast Quantum Gates for Neutral Atoms},
    author = {Jaksch, D. and Cirac, J. I. and Zoller, P. and Rolston, S. L. and C\^ot\'e, R. and Lukin, M. D.},
    journal = {Phys. Rev. Lett.},
    volume = {85},
    issue = {10},
    pages = {2208--2211},
    numpages = {0},
    year = {2000},
    month = {Sep},
    publisher = {American Physical Society},
    doi = {10.1103/PhysRevLett.85.2208},
    url = {https://link.aps.org/doi/10.1103/PhysRevLett.85.2208}
}

@article{Saffman2010,
    title = {Quantum information with {R}ydberg atoms},
    author = {Saffman, M. and Walker, T. G. and M\o{}lmer, K.},
    journal = {Rev. Mod. Phys.},
    volume = {82},
    issue = {3},
    pages = {2313--2363},
    numpages = {0},
    year = {2010},
    month = {Aug},
    publisher = {American Physical Society},
    doi = {10.1103/RevModPhys.82.2313},
    url = {https://link.aps.org/doi/10.1103/RevModPhys.82.2313}
}

@article{Levine2018,
    title = {High-Fidelity Control and Entanglement of {R}ydberg-Atom Qubits},
    author = {Levine, Harry and Keesling, Alexander and Omran, Ahmed and Bernien, Hannes and Schwartz, Sylvain and Zibrov, Alexander S. and Endres, Manuel and Greiner, Markus and Vuleti\ifmmode \acute{c}\else \'{c}\fi{}, Vladan and Lukin, Mikhail D.},
    journal = {Phys. Rev. Lett.},
    volume = {121},
    issue = {12},
    pages = {123603},
    numpages = {6},
    year = {2018},
    month = {Sep},
    publisher = {American Physical Society},
    doi = {10.1103/PhysRevLett.121.123603},
    url = {https://link.aps.org/doi/10.1103/PhysRevLett.121.123603}
}

@article{Evered2023,
    author={Evered, Simon J.
    and Bluvstein, Dolev
    and Kalinowski, Marcin
    and Ebadi, Sepehr
    and Manovitz, Tom
    and Zhou, Hengyun
    and Li, Sophie H.
    and Geim, Alexandra A.
    and Wang, Tout T.
    and Maskara, Nishad
    and Levine, Harry
    and Semeghini, Giulia
    and Greiner, Markus
    and Vuleti{\'{c}}, Vladan
    and Lukin, Mikhail D.},
    title={High-fidelity parallel entangling gates on a neutral-atom quantum computer},
    journal={Nature},
    year={2023},
    month={Oct},
    day={01},
    volume={622},
    number={7982},
    pages={268-272},
    issn={1476-4687},
    doi={10.1038/s41586-023-06481-y},
    url={https://doi.org/10.1038/s41586-023-06481-y}
}

@article{Euchner2025,
  title = {Rydberg atom arrays as quantum simulators for molecular dynamics},
  author = {Euchner, Simon and Lesanovsky, Igor},
  journal = {Phys. Rev. Res.},
  volume = {7},
  issue = {4},
  pages = {L042009},
  numpages = {7},
  year = {2025},
  month = {Oct},
  publisher = {American Physical Society},
  doi = {10.1103/r54t-myhc},
  url = {https://link.aps.org/doi/10.1103/r54t-myhc}
}

@article{Scholl2021,
    author={Scholl, Pascal
    and Schuler, Michael
    and Williams, Hannah J.
    and Eberharter, Alexander A.
    and Barredo, Daniel
    and Schymik, Kai-Niklas
    and Lienhard, Vincent
    and Henry, Louis-Paul
    and Lang, Thomas C.
    and Lahaye, Thierry
    and L{\"a}uchli, Andreas M.
    and Browaeys, Antoine},
    title={Quantum simulation of 2D antiferromagnets with hundreds of {R}ydberg atoms},
    journal={Nature},
    year={2021},
    month={Jul},
    day={01},
    volume={595},
    number={7866},
    pages={233-238},
    issn={1476-4687},
    doi={10.1038/s41586-021-03585-1},
    url={https://doi.org/10.1038/s41586-021-03585-1}
}

@article{Labuhn2016,
    author={Labuhn, Henning
    and Barredo, Daniel
    and Ravets, Sylvain
    and de L{\'e}s{\'e}leuc, Sylvain
    and Macr{\`i}, Tommaso
    and Lahaye, Thierry
    and Browaeys, Antoine},
    title={Tunable two-dimensional arrays of single {R}ydberg atoms for realizing quantum {I}sing models},
    journal={Nature},
    year={2016},
    month={Jun},
    day={01},
    volume={534},
    number={7609},
    pages={667-670},
    issn={1476-4687},
    doi={10.1038/nature18274},
    url={https://doi.org/10.1038/nature18274}
}

@article{Gross2017,
    author = {Christian Gross  and Immanuel Bloch},
    title = {Quantum simulations with ultracold atoms in optical lattices},
    journal = {Science},
    volume = {357},
    number = {6355},
    pages = {995-1001},
    year = {2017},
    doi = {10.1126/science.aal3837},
    URLX = {https://www.science.org/doi/abs/10.1126/science.aal3837},
    eprintX = {https://www.science.org/doi/pdf/10.1126/science.aal3837}
}

@article{Bloch2012,
    author={Bloch, Immanuel
    and Dalibard, Jean
    and Nascimb{\`e}ne, Sylvain},
    title={Quantum simulations with ultracold quantum gases},
    journal={Nature Physics},
    year={2012},
    month={Apr},
    day={01},
    volume={8},
    number={4},
    pages={267-276},
    issn={1745-2481},
    doi={10.1038/nphys2259},
    url={https://doi.org/10.1038/nphys2259}
}

@article{simons2021,
    title = {Rydberg atom-based sensors for radio-frequency electric field metrology, sensing, and communications},
    journal = {Measurement: Sensors},
    volume = {18},
    pages = {100273},
    year = {2021},
    issn = {2665-9174},
    doi = {https://doi.org/10.1016/j.measen.2021.100273},
    url = {https://www.sciencedirect.com/science/article/pii/S2665917421002361},
    author = {Matthew T. Simons and Alexandra B. Artusio-Glimpse and Amy K. Robinson and Nikunjkumar Prajapati and Christopher L. Holloway},
}

@article{Sedlacek2013,
    title = {Atom-Based Vector Microwave Electrometry Using {R}ubidium {R}ydberg Atoms in a Vapor Cell},
    author = {Sedlacek, J. A. and Schwettmann, A. and K\"ubler, H. and Shaffer, J. P.},
    journal = {Phys. Rev. Lett.},
    volume = {111},
    issue = {6},
    pages = {063001},
    numpages = {5},
    year = {2013},
    month = {Aug},
    publisher = {American Physical Society},
    doi = {10.1103/PhysRevLett.111.063001},
    url = {https://link.aps.org/doi/10.1103/PhysRevLett.111.063001}
}

@Article{Sedlacek2012,
    author={Sedlacek, Jonathon A.
    and Schwettmann, Arne
    and K{\"u}bler, Harald
    and L{\"o}w, Robert
    and Pfau, Tilman
    and Shaffer, James P.},
    title={Microwave electrometry with {R}ydberg atoms in a vapour cell using bright atomic resonances},
    journal={Nature Physics},
    year={2012},
    month={Nov},
    day={01},
    volume={8},
    number={11},
    pages={819-824},
    issn={1745-2481},
    doi={10.1038/nphys2423},
    url={https://doi.org/10.1038/nphys2423}
}

@article{Somaweera2025,
    AUTHOR = {Somaweera, Dinelka and Abdulghani, Amer and Odebowale, Ambali Alade and Berhe, Andergachew Mekonnen and Weerasinghe, Muthugalage I. U. and As’ham, Khalil and Al Ani, Ibrahim A. M. and Dumlao, Morphy C. and Miroshnichenko, Andrey E. and Hattori, Haroldo T.},
    TITLE = {Rydberg Atom-Based Sensors: Principles, Recent Advances, and Applications},
    JOURNAL = {Photonics},
    VOLUME = {12},
    YEAR = {2025},
    NUMBER = {12},
    ARTICLE-NUMBER = {1228},
    URL = {https://www.mdpi.com/2304-6732/12/12/1228},
    ISSN = {2304-6732},
    DOI = {10.3390/photonics12121228}
}

@article{Gorniaczyk2016,
    author={Gorniaczyk, H.
    and Tresp, C.
    and Bienias, P.
    and Paris-Mandoki, A.
    and Li, W.
    and Mirgorodskiy, I.
    and B{\"u}chler, H. P.
    and Lesanovsky, I.
    and Hofferberth, S.},
    title={Enhancement of {R}ydberg-mediated single-photon nonlinearities by electrically tuned {F}{\"o}rster resonances},
    journal={Nature Communications},
    year={2016},
    month={Aug},
    day={12},
    volume={7},
    number={1},
    pages={12480},
    issn={2041-1723},
    doi={10.1038/ncomms12480},
    url={https://doi.org/10.1038/ncomms12480}
}

@article{Nipper2012,
    title = {Highly Resolved Measurements of {S}tark-Tuned {F}\"orster Resonances between {R}ydberg Atoms},
    author = {Nipper, J. and Balewski, J. B. and Krupp, A. T. and Butscher, B. and L\"ow, R. and Pfau, T.},
    journal = {Phys. Rev. Lett.},
    volume = {108},
    issue = {11},
    pages = {113001},
    numpages = {5},
    year = {2012},
    month = {Mar},
    publisher = {American Physical Society},
    doi = {10.1103/PhysRevLett.108.113001},
    url = {https://link.aps.org/doi/10.1103/PhysRevLett.108.113001}
}

@article{Ravets2014,
    author={Ravets, Sylvain
    and Labuhn, Henning
    and Barredo, Daniel
    and B{\'e}guin, Lucas
    and Lahaye, Thierry
    and Browaeys, Antoine},
    title={Coherent dipole--dipole coupling between two single {R}ydberg atoms at an electrically-tuned {F}{\"o}rster resonance},
    journal={Nature Physics},
    year={2014},
    month={Dec},
    day={01},
    volume={10},
    number={12},
    pages={914-917},
    issn={1745-2481},
    doi={10.1038/nphys3119},
    url={https://doi.org/10.1038/nphys3119}
}

@incollection{Mokhberi2020,
    title = {Chapter Four - Trapped {R}ydberg ions: A new platform for quantum information processing},
    author = {Arezoo Mokhberi and Markus Hennrich and Ferdinand Schmidt-Kaler},
    booktitle = {Advances In Atomic, Molecular, and Optical Physics},
    publisher = {Academic Press},
    volume = {69},
    pages = {233-306},
    year = {2020},
    issn = {1049-250X},
    doi = {https://doi.org/10.1016/bs.aamop.2020.04.004},
    url = {https://www.sciencedirect.com/science/article/pii/S1049250X20300045},
}

@article{Higgins2021,
  title = {Observation of second- and higher-order electric quadrupole interactions with an atomic ion},
  author = {Higgins, Gerard and Zhang, Chi and Pokorny, Fabian and Parke, Harry and Jansson, Erik and Salim, Shalina and Hennrich, Markus},
  journal = {Phys. Rev. Res.},
  volume = {3},
  issue = {3},
  pages = {L032032},
  numpages = {6},
  year = {2021},
  month = {Aug},
  publisher = {American Physical Society},
  doi = {10.1103/PhysRevResearch.3.L032032},
  url = {https://link.aps.org/doi/10.1103/PhysRevResearch.3.L032032}
}

@misc{bao2025,
      title={Microwave-Dressing of {R}ydberg States in a Trapped {C}alcium Ion}, 
      author={Han Bao and Alexander Schulze-Makuch and Ferdinand Schmidt-Kaler},
      year={2025},
      eprint={2504.21241},
      archivePrefix={arXiv},
      primaryClass={physics.atom-ph},
      url={https://arxiv.org/abs/2504.21241}, 
}

@phdthesis{pokorny_thesis,
    author = {Fabian Pokorny},
    title = {A microwave dressed {R}ydberg ion},
    school = {Stockholm university},
    year = {2020}
}

@article{Higgins2019,
    title = {Highly Polarizable {R}ydberg Ion in a {P}aul Trap},
    author = {Higgins, Gerard and Pokorny, Fabian and Zhang, Chi and Hennrich, Markus},
    journal = {Phys. Rev. Lett.},
    volume = {123},
    issue = {15},
    pages = {153602},
    numpages = {6},
    year = {2019},
    month = {Oct},
    publisher = {American Physical Society},
    doi = {10.1103/PhysRevLett.123.153602},
    url = {https://link.aps.org/doi/10.1103/PhysRevLett.123.153602}
}

@article{Zhang2020,
    author={Zhang, Chi
    and Pokorny, Fabian
    and Li, Weibin
    and Higgins, Gerard
    and P{\"o}schl, Andreas
    and Lesanovsky, Igor
    and Hennrich, Markus},
    title={Submicrosecond entangling gate between trapped ions via {R}ydberg interaction},
    journal={Nature},
    year={2020},
    month={Apr},
    day={01},
    volume={580},
    number={7803},
    pages={345-349},
    issn={1476-4687},
    doi={10.1038/s41586-020-2152-9},
    url={https://doi.org/10.1038/s41586-020-2152-9}
}

@article{qutip,
    title = {QuTiP 2: A {P}ython framework for the dynamics of open quantum systems},
    journal = {Computer Physics Communications},
    volume = {184},
    number = {4},
    pages = {1234-1240},
    year = {2013},
    issn = {0010-4655},
    doi = {https://doi.org/10.1016/j.cpc.2012.11.019},
    url = {https://www.sciencedirect.com/science/article/pii/S0010465512003955},
    author = {J.R. Johansson and P.D. Nation and Franco Nori}
}

@misc{iminuit,
    author={Hans Dembinski and Piti Ongmongkolkul et al.},
    title={scikit-hep/iminuit},
    DOI={10.5281/zenodo.3949207},
    publisher={Zenodo},
    year={2020},
    month={Dec},
    url={https://doi.org/10.5281/zenodo.3949207}
}

@article{James:1975dr,
    author = "James, F. and Roos, M.",
    title = "{Minuit: A System for Function Minimization and Analysis of the Parameter Errors and Correlations}",
    reportNumber = "CERN-DD-75-20",
    doi = "10.1016/0010-4655(75)90039-9",
    journal = "Comput. Phys. Commun.",
    volume = "10",
    pages = "343--367",
    year = "1975"
}

\clearpage
\appendix

\section{Appendix}
\subsection{Microwave setup}\label{appendix:MWcalibration}
\begin{figure}[!t]
    \centering
    \includegraphics{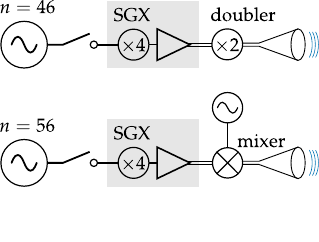}
    \vspace{-1cm}
    \caption{MW generation for both principal quantum numbers $n=46$ (top) and $n=56$ (bottom). A seed signal is switched before being quadrupled in frequency and amplified in a signal generator extension (SGX) module (azure background). For $n=46$, the signal, in the WR15 frequency range, is once more frequency doubled, resulting in a signal in the WR8 frequency range. For $n=56$, the signal after the SGX is upconverted with an RF signal of $\SIrange{200}{400}{\mega\hertz}$. In both cases, the signal is coupled to free space with a conical horn antenna. Single lines indicate coaxial connections; double lines indicate waveguides.}
    \label{fig:MWsetup}
\end{figure}

A signal generator with a maximum frequency of \SI{20}{\giga\hertz} provides the seed signal. This signal is switched using a solid-state switch before being fed into a signal generator extension (SGX) module, quadrupling and amplifying the signal. For the experiments with $n=46$, an additional frequency doubler allows us to reach the required \SI{120}{\giga\hertz} range, see also Fig.~\ref{fig:MWsetup} (top).

\begin{figure}[!t]
    \centering
    \includegraphics{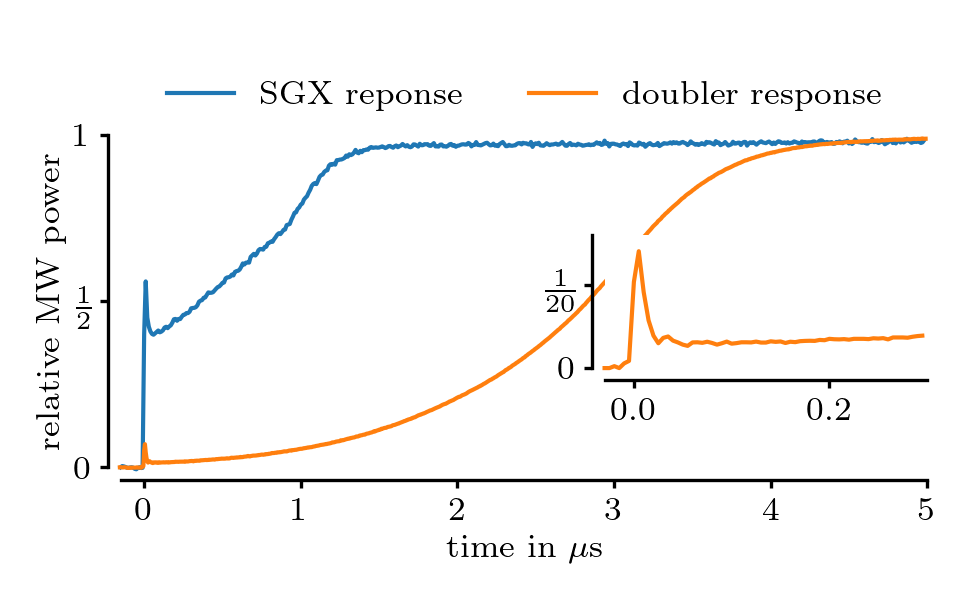}
    \caption{Switching of the MW at different points in the signal chain. While the seed signal is switched on \si{\nano\second} timescales, the SGX output initially rises to half power, with a slow, linear rise to full power within \SI{1.5}{\micro\second}. The MW power after the doubler rises very slowly and reaches full power only after \SI{4}{\micro\second}, however there is an initial peak and a low, but non-zero, linear rise of MW power for the first $100$s of \si{\nano\second} (see insert).}
    \label{fig:MWswitching}
\end{figure}

The switch itself has a rise and fall time of \SI{1}{\nano\second}, however we find a much slower response of the final MW radiation sent to the ion, caused by the slow response of the SGX module and the last frequency doubler. Figure \ref{fig:MWswitching} shows the response of MW power at different points in the signal chain. The output power features an initial peak on \si{\nano\second} timescales after switching on, followed by a low but non-zero linear rise for the first 100s of \si{\nano\second}. The peak can either be caused by an initial peak in MW power or by an artifact of the MW detector. We therefore add an initial pulse area to the MW pulse in the simulation and fit its value to the offset of the MW Rabi oscillations presented in Figure~\ref{fig:pulse_sequence_MWrabi} of the main text.

For the MW sweeping experiments, a principal quantum number of $n=56$ was used. To be able to sweep the MW frequency, the last frequency doubler is replaced with a balanced upconverter (Fig.~\ref{fig:MWsetup}, bottom), reducing the frequency that can be reached to the WR15 frequency band. By choosing a higher Rydberg state, the splitting between the Rydberg $S$ and $P$ states becomes smaller, allowing us to address the $rS \leftrightarrow rP$ transition with the reduced MW frequency. We use a commercial DDS (Sinara 4412 DDS Urukul) with fast frequency control to mix an RF signal of \mbox{\SIrange{200}{400}{\mega\hertz}} onto the MW. This RF signal is then swept in frequency to create the frequency sweep of the MW radiation.

\subsection{Simulation of the Microwave Rabi oscillations}
\begin{table*}
    \centering
    \begin{tabular}{lrrrr}
        Parameter & opt. value (MW Rabi) & cal. value (MW Rabi) & opt. value (MW sweep) & cal. value (MW sweep)\\
        \hline \hline
        effective Rydberg Rabi frequency $\Omega_\mathrm{R}$ & $2\pi \times \SI{3.334(2)}{\mega\hertz}$ & $2\pi \times \SI{3.6(2)}{\mega\hertz}$ 
        & $2\pi \times \SI{2.37(2)}{\mega\hertz}$ & $2\pi \times \SI{2.4(2)}{\mega\hertz}$\\
        
        Detuning $\Delta_{305}$ from ${rS}$ or $\ket{+}_\mathrm{0pol}$ & $-2\pi \times \SI{2.39(12)}{\mega\hertz}$ & $-2\pi \times \SI{2.0(4)}{\mega\hertz}$ 
        & $2\pi \times \SI{1.6(2)}{\mega\hertz}$ & $-2\pi \times \SI{0.9(4)}{\mega\hertz}$\\
        
        Dephasing $\gamma_{243/305}$ of ${S}$, ${P}$ and ${rS}$ & $2\pi \times \SI{0.11(4)}{\mega\hertz}$ & -
        & - & $2\pi \times \SI{0.11(4)}{\mega\hertz}^*$\\
        
        \hline
        
        MW Rabi frequency $\Omega_\mathrm{MW}$ & $2\pi \times \SI{11.1(3)}{\mega\hertz}$ & $2\pi \times \SI{12.3(5)}{\mega\hertz}$ 
        & - & $2\pi \times \SI{65}{\mega\hertz}$ \\
        
        MW detuning $\Delta_\mathrm{MW}$ & $-2\pi \times \SI{2.68(13)}{\mega\hertz}$ & \SI{0}{\mega\hertz} 
        & - & $2\pi \times \SIrange{0}{50}{\mega\hertz}$\\
        
        Dephasing $\gamma_\mathrm{R}$ of ${rS}$ and ${rP}$ & $2\pi \times \SI{0.16(2)}{\mega\hertz}$ & - 
        & - & $2\pi \times \SI{0.02(6)}{\mega\hertz}^\dagger$ \\
        
        Initial MW pulse area & \SI{11.07(13)}{\nano\second} & - & - & -\\
    \end{tabular}
    \caption{Overview of the optimized and calibrated (where available) values for the simulation. The first three values were fitted to the Rabi oscillations to the Rydberg state ${rS}$ for the MW Rabi experiments, or $\ket{+}$ for the MW sweeping experiments. These were kept constant for the simulation of the MW Rabi oscillations, only the last four values were optimized to represent the experimental data. All parameters are within $2\sigma$ of the calibrated value, except for the MW detuning $\Delta_\mathrm{MW}$, which most likely stems from a wrong calibration or off-resonant coupling to other levels leading to a Stark shift of one of the Rydberg levels. $^*$This value is taken from the optimization of the MW Rabi frequency and is fixed for the simulation of the MW sweep experiments. $^\dagger$This value was fittend together with the first values to the Rabi oscillations between $D$ and $\ket{+}_\mathrm{0pol}$.}
    \label{tab:sim_fitparams}
\end{table*}

The simulated dynamics were obtained by numerically solving the Lindblad master equation
\begin{multline}
    \Dot{\rho}(t) = - \frac{1}{2} \left[ H(t), \rho(t) \right] \\
    + \sum_k \frac{1}{2} \left( 2 C_k \rho(t) C_k^\dag - \rho(t) C_k^\dag C_k - C_k^\dag C_k \rho(t) \right)
\end{multline}
with the system Hamiltonian $H$, density matrix $\rho$ and collapse operators $C_k$ which describe the coupling to the environment, in our case spontaneous decay and dephasing. The integration of the Lindblad master equation was done in Python using the QuTIP package \cite{qutip}. The Hamiltonian consists of the Rydberg excitation and MW transitions \cite{wilkinson2024}:

\begin{widetext}
\begin{align}
    H =& \frac{\hbar\Omega_{243}(t)}{2} \ket{P}\bra{D} + \frac{\hbar}{2} \Delta_{243} \ket{P}\bra{P} 
    + \frac{\hbar\Omega_{305}(t)}{2} \ket{rS}\bra{P} + \frac{\hbar}{2} \left(\Delta_{243} + \Delta_{305}\right) \ket{rS}\bra{rS} \nonumber\\
    &+ \frac{\hbar\Omega_\mathrm{MW}(t)}{2} \ket{rP}\bra{rS} + \frac{\hbar}{2}\left(\Delta_{243} + \Delta_{305} + \Delta_\mathrm{MW}(t)\right) \ket{rP}\bra{rP} 
    + h.c.
\end{align}
\end{widetext}
With time dependent Rabi frequencies of the individual transitions $\Omega_{243/305/\mathrm{MW}}$ and detunings $\Delta_{243/305/\mathrm{MW}}$, where the detuning of the MW is time-dependent for the adiabatic transfer experiments.

As collapse operators, we consider spontaneous decay
\begin{equation}
    C_{i} = \frac{1}{\sqrt{2\tau_i}} \ket{S}\bra{i}
\end{equation}
from $\ket{i} = \ket{P}, \ket{rS}, \ket{rP}$ to $\ket{S}$, with lifetimes of \mbox{$\tau_{{P}} = \SI{37}{\nano\second}$} \cite{Kiruga2026}, $\tau_{{rS}}^{46} = \SI{4.6}{\micro\second}$, $\tau_{{rP}}^{46} = \SI{26.6}{\micro\second}$, $\tau_{{rS}}^{56} = \SI{8.2}{\micro\second}$ and \mbox{$\tau_{{rP}}^{56} = \SI{40.4}{\micro\second}$} at room temperature (\SI{300}{\kelvin}) \cite{Weibin}. Furthermore, we consider dephasing
\begin{align}
    C_{243} &= \sqrt{\gamma_{243}} \left( \ket{P}\bra{P} - \ket{D}\bra{D} \right) \nonumber\\
    C_{305} &= \sqrt{\gamma_{305}} \left( \ket{rS}\bra{rS} - \ket{P}\bra{P} \right) \nonumber\\
    C_\mathrm{R} &= \sqrt{\gamma_\mathrm{R}} \left( \ket{rP}\bra{rP} - \ket{rS}\bra{rS} \right)
\end{align}
due to the Rydberg excitation laser linewidths ($C_{243/305}$), affecting states ${D}$, ${P}$ and ${rS}$. We also include dephasing of the Rydberg states ${rS}$ and ${rP}$ due to fluctuating state energies ($C_\mathrm{R}$), which is relevant for Rydberg states that are highly sensitive to electric field noise and fluctuate in energy with electric field fluctuations \cite{pokorny2020, Higgins2021}.

Rabi frequencies were calibrated via the Autler-Townes scheme described in the main text and \cite{Mokhberi2020, Higgins2021, bao2025}. For the simulation of the Rabi oscillations to the ${rS}$ state presented in Figure~\ref{fig:levels_and_UV_Rabi} b), values fitted to the oscillations observed were:
\begin{itemize}
    \item \emph{Rydberg excitation Rabi frequency $\Omega_\mathrm{R}$:} a correction to the Rabi frequency from the calibration, the same factor was applied for both Rydberg excitation lasers.
    \item \emph{Detuning} $\Delta_{305}$: due to off-resonant coupling to other levels that is not captured in the simulation, we chose to keep this as a free parameter.
    \item \emph{Decoherence of the ${S}$, ${P}$ and ${rS}$ levels} due to Rydberg excitation laser linewidths.
\end{itemize}

\begin{table}[b]
    \centering
    \begin{tabular}{l r}
        Error source & reduction in efficiency \\
        \hline\hline
        Rydberg excitation laser linewidths & \SI{0.4}{\percent} \\
        Rydberg state dephasing & \SI{2.2}{\percent} \\
        Rydberg excitation detuning & \SI{0.005}{\percent} \\
        MW detuning & \SI{7.4}{\percent} \\
        \hline
        Total inefficiency & \SI{8.5(5)}{\percent}
    \end{tabular}
    \caption{Error budget of the MW $\pi$-pulse. The biggest contribution is from the detuning of the MW from resonance. Furthermore, Rydberg state dephasing plays a significant role.}
    \label{tab:MWerrorbudget}
\end{table}

For the MW Rabi oscillations presented in Figure \ref{fig:pulse_sequence_MWrabi} b), the parameters above were fixed at their optimum values and the following parameters were fitted to the experimental data:
\begin{itemize}
    \item \emph{MW Rabi frequency $\Omega_\mathrm{MW}$}: a correction from the calibrated value, to capture drifts in the MW power and systematic errors in the calibration process.
    \item \emph{MW detuning} $\Delta_\mathrm{MW}$: Due to an asymmetric and relatively broad MW resonance, we chose to include this as a free parameter. Furthermore, off-resonant coupling to other levels can introduce a Stark shift not captured in the simulation.
    \item \emph{Dephasing} $\gamma_\mathrm{R}$ due to fluctuations in energy of the ${rS}$ and ${rP}$ levels.
    \item \emph{Initial MW pulse area} to capture the initial peak in MW power (see also Fig. \ref{fig:MWswitching}).
\end{itemize}
All fitted parameters are summarized in Table~\ref{tab:sim_fitparams}, together with their calibrated values where available.
The optimization was done with Python using the iminuit package \cite{iminuit, James:1975dr}.

The efficiency of the MW $\pi$-pulse was estimated by repeated simulation of the state populations while drawing random numbers for the fit values according to their optimum values and covariance matrices leading to a total inefficiency of \SI{8.5(5)}{\percent}. Furthermore, we investigated different sources of inefficiency (see Tab.~\ref{tab:MWerrorbudget}) by assuming a perfectly calibrated experiment and introducing individual imperfections in the simulation. We find that the MW detuning has the biggest impact on the efficiency, providing an inefficiency of $\SI{7.3}{\percent}$, followed by dephasing of the Rydberg levels due to energy fluctuations of \SI{2.2}{\percent}. The remaining error sources investigated are all well below \SI{1}{\percent}.

\begin{figure}
    \centering
    \includegraphics{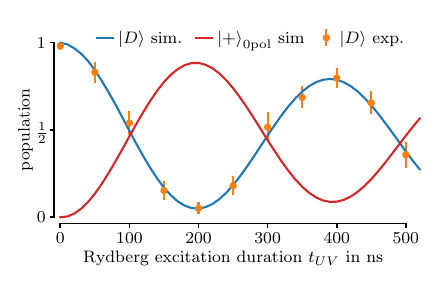}
    \caption{Rabi oscillations between ${D}$ and the zero polarizability Rydberg state $\ket{+}_\mathrm{0pol}$. Solid lines show simulated dynamics and dots show experimentally recorded populations. We achieve an excitation efficiency of \SI{88}{\percent}. Error bars represent $1\sigma$ confidence intervals and are due to quantum projection noise.}
    \label{fig:Rabi_osc_0pol}
\end{figure}

\subsection{Simulation of the Microwave sweep}
The calibration of experimental values for the MW sweeping experiments presented was done in the same way as for the MW Rabi oscillations. 
First, Rabi oscillations to the zero polarizability state $\ket{+}_\mathrm{0pol}$ were measured (see Fig.\,\ref{fig:Rabi_osc_0pol}), followed by the sweep experiments presented in the main text. The Rabi oscillations were analyzed and simulated in the same manner as for the results presented in Figure \ref{fig:levels_and_UV_Rabi} b), the fitted and calibrated values are presented in Table~\ref{tab:sim_fitparams}. 
Here, we assume the dephasing from the Rydberg excitation laser linewidths and the initial pulse area of the Rydberg excitation lasers not to have changed and did not optimize them again. Since the experiments involve MW-dressed Rydberg states with continuous MW-dressing, the initial MW pulse area does not play a role, and small deviations from the calibrated MW Rabi frequency $\Omega_\mathrm{MW}$ and detuning $\Delta_\mathrm{MW}$ only lead to small changes in the detuning $\Delta_\mathrm{R}$ of the dressed state $\ket{+}$ from the bare Rydberg state ${rS}$. Experimentally we verify and adjust these small changes by probing the resonance of the dressed state with the Rydberg excitation lasers, and we neglect these shifts in the simulation. However, the Rydberg state dephasing $\gamma_\mathrm{R}$ depends strongly on the principal quantum number $n$ and the electric field noise, which is why we also fitted this value to Rabi oscillations to the $\ket{+}_\mathrm{0pol}$ state.

\clearpage

\end{document}